\documentclass[a4paper]{jpconf}
\usepackage{graphicx}
\usepackage{amssymb}

\begin{document}

\title{Microscopic Structural and Dynamical Properties of Amorphous Metallic Alloy $Ni_{33}Zr_{67}$ at the Temperature $T=300K$}
\author{Ramil M. Khusnutdinoff, Anatolii V. Mokshin and Ilgiz I. Khadeev}

\address{Department of Computational Physics, \\Institute of Physics, Kazan Federal University,\\
Kremlevskaya Street 18, 420008 Kazan, Russia}

\ead{r.khusnutdinoff@gmail.com}

\begin{abstract}
We study the structural properties and the collective microscopic dynamics of atoms in the amorphous metallic alloy $Ni_{33}Zr_{67}$ at the temperature $T=300K$ by molecular dynamics simulations. The calculated equilibrium structural and dynamical characteristics are compared with the experimental data on neutron diffraction and on inelastic X-ray scattering. We present the interpretation of observed structural relaxation of the microscopic density fluctuations of particles for amorphous metallic alloy in the framework of the recurrent relation approach. The results of theoretical calculations of the intensity of scattering $I(k,\omega)$ for amorphous $Ni_{33}Zr_{67}$ are in a good agreement with the results of computer simulation as well as with the experimental data on inelastic X-ray scattering.
\end{abstract}



\section{Introduction}

Design and production of new materials are important fields in the continuous development of science and technology. Bulk metallic glasses (BMG) belong to the advanced materials that attract the special interest of scientists and technologists because of their unique physicochemical and mechanical properties \cite{Clement1960,Greer1995,Johnson1999}. For example, amorphous alloys containing transition metal ($Fe, Co, Ni$) as a main component have high tensile strength, which exceeds by more than twice of their crystalline counterparts. Some of these amorphous alloys also have a high corrosion resistance, the excellent magnetic and electrical properties. The unique properties of the BMG are related with the local structural disorder in the atomic components of the mixture \cite{Khusnutdinoff2010}.
Amorphous metal alloys $Ni_{x}Zr_{100-x}$ are characterized by the following interesting features \cite{Omoto1998}: (i) the structure of the alloys has strong short-range topological order, which significantly affects on the dispersion curve in the amorphous state; (ii) the alloys have a large coherent neutron scattering cross section, which is an important feature at the comparison of the simulation results and the theoretical calculations with the experimental data on neutron diffraction. In addition, some metal alloys (such as the known,  $Zr_{41.2}Ti_{13.8}Cu_{12.5}Ni_{10.0}Be_{22.5}$ - Vitreloy 1) form the bulk metallic glass phase at the critical cooling rates of $\sim 1~K/sec$ \cite{Johnson2002}. These features have stimulated many experimental and theoretical studies. So, for example, the influence of the concentration dependence of the $Ni_xZr_{100-x}$ alloys ($x = 5, 10, 16.7, 33.3, 0.5, 66.7, 83.3, 90.0, 95.0$) on the processes of crystallization of the bulk metallic glass was investigated by using the computer simulation methods in Ref. \cite{Yang2007}. The authors demonstrated that the highest degree of crystallization of the alloy is related with high nickel concentration in the system (in particular, $Ni_{66.7}Zr_{33.3}$-alloy), where crystallization in the alloys is related with increasing of the number of quasicrystalline icosahedral clusters.
Further, the study of the dynamic properties of the $Ni_{33}Zr_{67}$-alloy in amorphous and crystalline phase was performed in Ref. \cite{Omoto2002} by using the inelastic neutron scattering experiment. By comparing the spectra of the dynamic structure factor, $S(k,\omega)$, in amorphous and crystalline phases, the authors have concluded that the origin of the collective excitations in  $Ni_{33}Zr_{67}$ metallic glass is closely associated with the optical excitations which observed in the crystalline phase. In addition, the three optical modes, which exist in the crystalline  $Ni_{33}Zr_{67}$-alloy, can also be observed on the short time scales in the amorphous phase. However, the recent experimental studies have questioned this conclusion \cite{Scopigno2006}.

In this paper we investigate the microscopical structural and dynamical properties of amorphous  $Ni_{33}Zr_{67}$ metallic alloy at the temperature $T=300K$ to elucidate the mechanism of propagation of collective excitations.

\section{Details of simulation}

\begin{figure}[htb]
\begin{center}
\includegraphics[width=16cm]{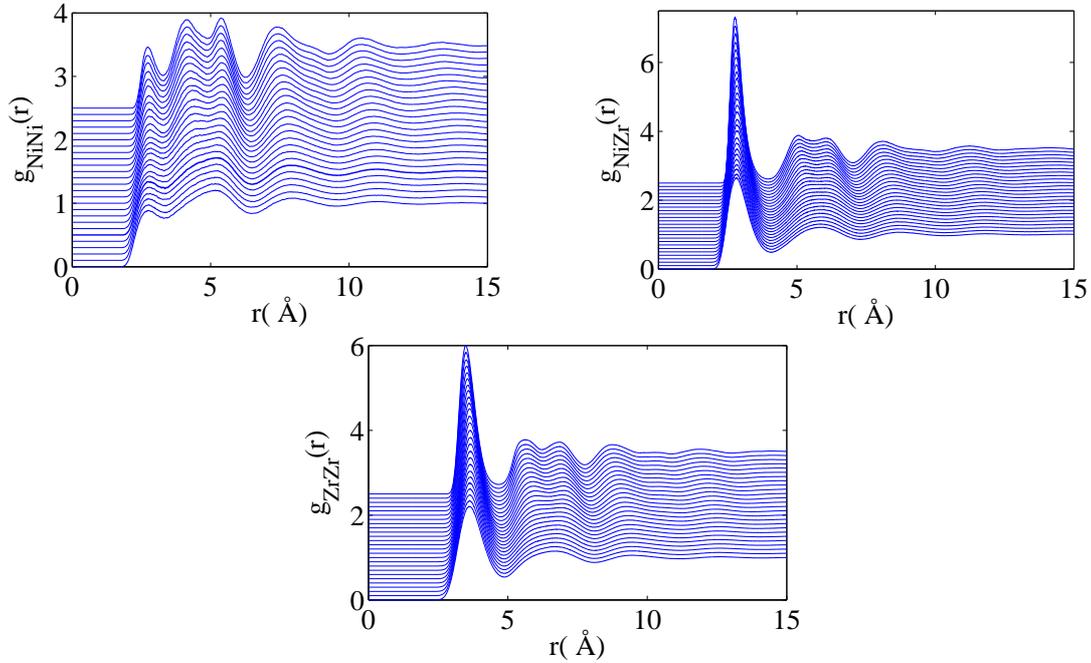}
\end{center}
\caption{(Color online) Partial radial distribution functions of particles for $Ni_{33}Zr_{67}$ metallic alloy at the temperatures $T$ from $300K$ (top curve) to $3000K$ (bottom curve).}
\label{fig_RDF}
\end{figure}

The molecular dynamics (MD) simulation of the amorphous metallic alloy $Ni_{33}Zr_{67}$ was performed at the temperature $T=300K$. The investigated system was consisted of  $N=N_{Ni}+N_{Zr}=10976$  particles, where $N_{Ni}=3535$ and $N_{Zr}=7441$; the numerical density was $n=0.0427 \textrm{\AA}^{-3}$. The interaction of atoms was carried out with the potential of interparticle interaction \cite{Hausleitner1990}
\begin{equation}
U_{\alpha,\beta}(r)=A_{\alpha,\beta}\Bigg[\frac{1}{\big(\zeta_{\alpha,\beta} r-a_{\alpha,\beta}\big)^n} -1\Bigg]\exp{\bigg(\frac{1}{\zeta_{\alpha,\beta} r-b_{\alpha,\beta}} \bigg)}, \ \ \ \ \bigg(r<\frac{a_{\alpha,\beta}}{\zeta_{\alpha,\beta}}\bigg).
\label{Eq_Pot}
\end{equation}
Here $\alpha,\beta \in \{Ni,Zr\}$ and $A_{\alpha,\beta}$, $\zeta_{\alpha,\beta}$, $a_{\alpha,\beta}$, $b_{\alpha,\beta}$ -- are the potential parameters, taken from Ref. \cite{Hausleitner1990}. All simulations are carried out in the isobaric-isothermal ensemble ($NpT$). To maintain the system in the thermal equilibrium, the  Berendsen's thermostat and barostat were applied. The amorphous alloy at the temperature $T=300K$ was obtained by a rapid cooling from the liquid state at the temperature $T=3000K$ with the rate $dT/dt=10^{13}~K/sec$. The integration of the equations of motion of the particles was performed using the velocity Verlet algorithm with the timestep $10^{-15}$ sec.

\begin{figure}[htb]
\begin{center}
\includegraphics[width=15cm]{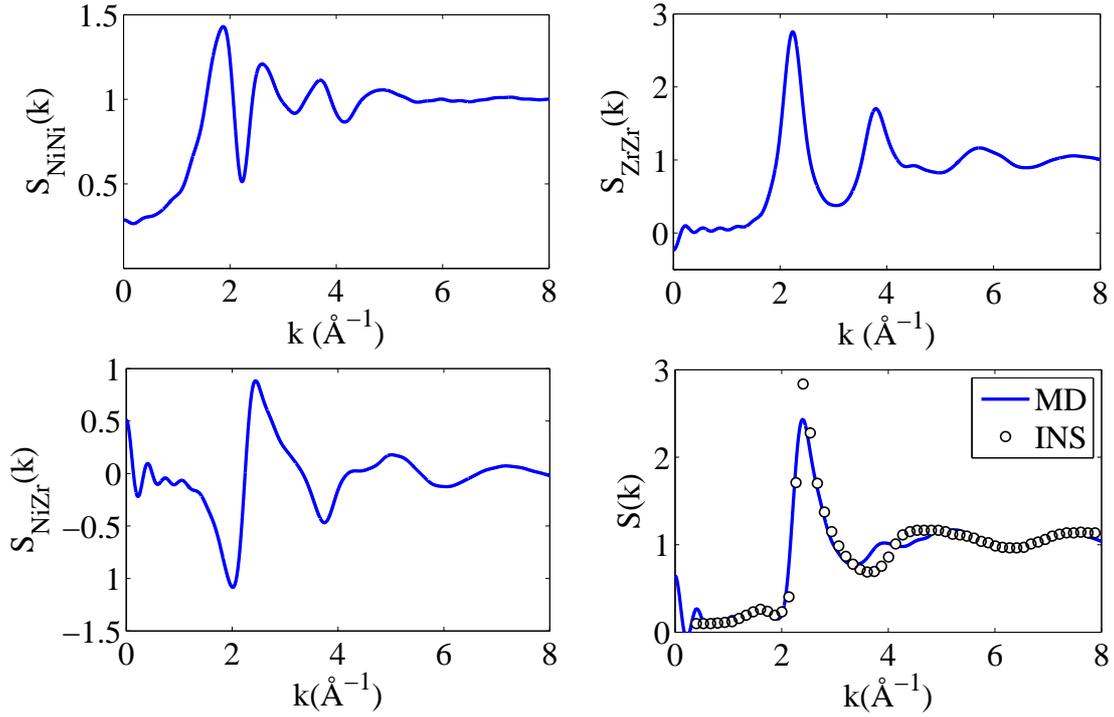}
\end{center}
\caption{(Color online) Static structure factor of $Ni_{33}Zr_{67}$-alloy at the temperature $T=300K$.}
\label{fig_Sk}
\end{figure}

\section{Simulation results and discussion}
The structural features of metallic alloy were analyzed by means of the partial radial distribution function:
\begin{equation}
g_{\alpha,\beta}(r)=\frac{V}{N_{\alpha}N_{\beta}}\Bigg\langle \sum_{i=1}^{\alpha}
\sum_{j=1}^{\beta}\delta(r-r_{ij})  \Bigg\rangle, ~~~~~~~~ \alpha,\beta \in \{Ni,Zr\}
\label{Eq_Pcf}
\end{equation}
and the partial static structure factor \cite{Allen_Tildesley}:
\begin{equation}
S_{\alpha,\beta}(k)=\delta_{\alpha,\beta}+4\pi n\int_0^{\infty}r^2\bigg[ g_{\alpha,\beta}(r)-1 \bigg]\frac{\sin(kr)}{kr}dr.
\label{Eq_PSk}
\end{equation}
Here, $k$ is the wave-number. The total static structure factor was found by means of the spatial Fourier transformation of the total radial distribution function \cite{Andonov1985}:
\begin{equation}
g(r)=\frac{1}{(c_1b_1+c_2b_2)^2}\sum_{\alpha,\beta}^2\sqrt{c_{\alpha}c_{\beta}}f_{\alpha}f_{\beta} g_{\alpha,\beta}(r), ~~~~~~~~ \alpha,\beta \in \{Ni,Zr\},
\label{Eq_RDF}
\end{equation}
where, $c$ is the concentration of particles, $f$ is the scattering length (in the case of neutron scattering).
Fig. 1 shows the temperature dependence of the particle distribution functions, $g_{\alpha,\beta}(r)$, calculated on the basis of molecular dynamics simulation. The figure demonstrates that the $Ni_{33}Zr_{67}$-system is completely transformed into an amorphous state with a cooling: the distribution function is characterized a pronounced first peak and a splitting of the second one \cite{Mokshin2006a}.
Fig. 2 shows the partial static structure factor obtained on the basis of computer simulation while the comparison of the MD results for $S(k)$ with the experimental data on neutron diffraction \cite{Lee1982} is presented in Fig 2(d). It can be seen from the figure that the chosen model of interatomic interaction potential \cite{Hausleitner1990} reproduces in detail the structure of amorphous alloy.

\begin{figure}[htb]
\begin{center}
\includegraphics[width=17cm]{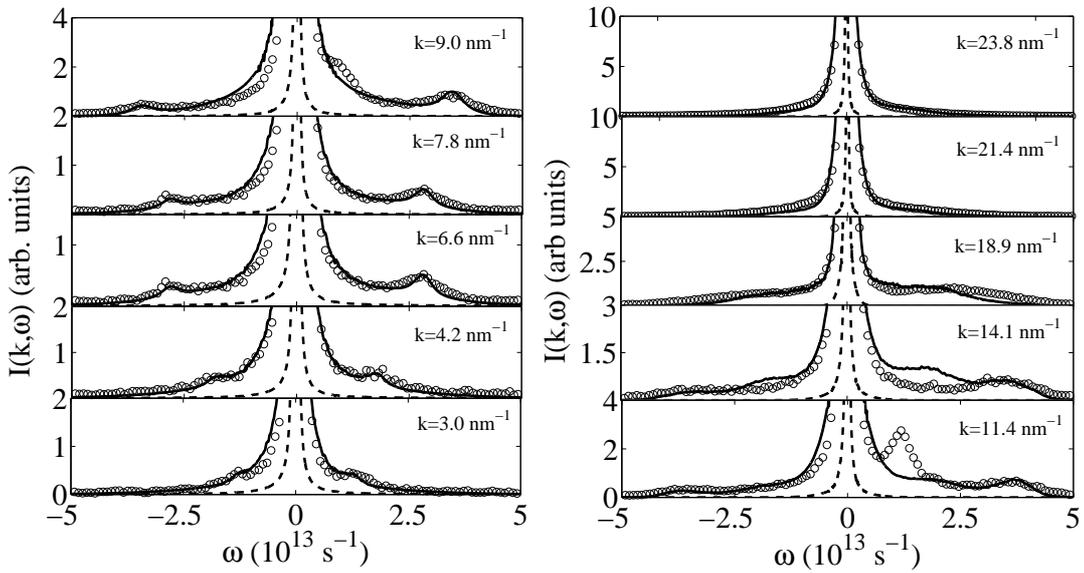}
\end{center}
\caption{Intensity of scattering for the metallic alloy $Ni_{33}Zr_{67}$ at the temperature $T=300K$. The solid line represents the simulation results, taking into account the detailed balance condition (6) and the experimental resolution function (7); $(\circ ~ \circ ~ \circ)$ -- the experimental data on inelastic X-ray scattering \cite{Scopigno2006}. The dotted line shows the experimental resolution function $R(k,\omega)$.}
\label{fig_Ikw01}
\end{figure}

\begin{figure}[htb]
\includegraphics[width=16cm]{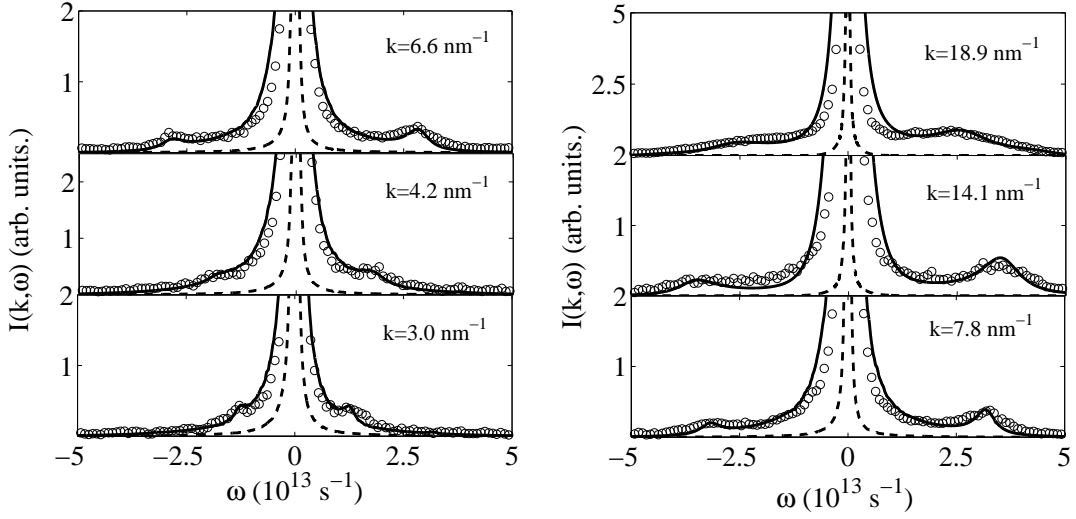}
\caption{Intensity of scattering for metallic alloy $Ni_{33}Zr_{67}$ at the temperature $T=300K$. The solid line represents the results of theoretical calculations, taking into account the detailed balance condition (6) and the experimental resolution function (7); $(\circ ~ \circ ~ \circ)$ -- the experimental data on inelastic X-ray scattering \cite{Scopigno2006}. The dotted line shows the experimental resolution function $R(k,\omega)$.}
\label{fig_Ikw02}
\end{figure}
The microscopic dynamics of the amorphous metal alloy was analyzed by means of the dynamic structure factor $S(k,\omega)$. To compare the simulation results with the experimental data, we have calculated the intensity of the inelastic X-ray scattering, $I(k,\omega)$, which is related with the dynamic structure factor by the following relations \cite{Khusnutdinoff2009}:
\begin{eqnarray}
I(k,\omega)=E(k)\int R(k,\omega-\omega ') S_q(k,\omega ')d\omega ', \\
S_q(k,\omega)=\frac{\hbar \beta \omega}{1-\mathrm{e}^{-\hbar
\beta \omega}} S(k,\omega). \label{Eq_Intens}
\end{eqnarray}
Here $S_q(k,\omega)$ is the quantum dynamic structure factor, $\beta=1/(k_BT)$ is the inverse temperature, $E(k)$
is a normalization factor, and $R(k,\omega)$ is the experimental resolution function
\begin{eqnarray}
R(k,\omega)=\frac{1}{\sqrt{\pi}\omega_0(k)}\mathrm{e}^{-\omega^2/\omega_0(k)^2}
\label{res_fun}
\end{eqnarray}
which satisfies the normalization condition
\begin{eqnarray}
\int_{-\infty}^{\infty}R(k,\omega)d\omega=1. \label{res_norm}
\end{eqnarray}
Comparison of the simulation results for the scattering intensity $I(k,\omega)$ with the experimental data on inelastic X-ray scattering is given in Fig. 3. As can be seen from the figure, the simulation results reproduce accurately the high frequency features of the X-ray scattering intensity in a wide range of wave numbers: they predict the positions, heights, and decrease of the side peaks. However, the simulation results do not reveal the experimentally-observed low-frequency excitations in  $I(k,\omega)$ at the values of  $k=9\div 11.4 ~nm^{-1}$.

Accordingly to the theoretical approach developed \cite{Mokshin2007,Yulmetyev2002} within Lee's recurrent relation formalism \cite{Lee1983a,Lee1983b} the dynamic structure factor can be taken in the form \cite{Mokshin2006b}:
\begin{eqnarray}
\lefteqn{
S(k,\omega)=\frac{S(k)}{2\pi}\Delta_1(k)\Delta_2(k)\Delta_3(k)
\sqrt{4\Delta_4(k)-\omega^{2}}
\Bigg\{\Delta_1^2(k)\Delta_3^2(k)
+\omega^{2}\Bigg[-2\Delta_1(k)\Delta_3^2(k)+}
\nonumber\\&&~~
+\Delta_1(k)^2\Delta_4(k)-\Delta_1^2(k)\Delta_3(k)+2\Delta_1(k)
\Delta_2(k)\Delta_4(k)-\Delta_1(k)\Delta_2(k)\Delta_3(k)+\nonumber\\&&~~ +\Delta_2^2(k)\Delta_4(k)\Bigg]+
\omega^{4}\Bigg[\Delta_3^2(k)-2\Delta_1(k)\Delta_4(k)+2\Delta_1(k)
\Delta_3(k)-2\Delta_2(k)
\Delta_4(k)+ \nonumber\\&&~~+\Delta_2(k)\Delta_3(k)\Bigg]
+\omega^{6}\Bigg[\Delta_4(k)-\Delta_3(k)\Bigg]\Bigg\}^{-1}.
\label{Eq_14}
\end{eqnarray}
Here, $S(k)$ is the static structure factor; $\Delta_1(k)$, $\Delta_2(k)$, $\Delta_3(k)$ and $\Delta_4(k)$ are the frequency relaxation parameters:
\begin{eqnarray}
 \label{fr_c}
\Delta_{1}(k)&=&\frac{k_BT}{m}\frac{k^2}{S(k)}, \nonumber\\
\Delta_{2}(k)&=&\frac{k_BT}{m}k^2  \left (
3-\frac{1}{S(k)} \right ) + \frac{n}{m} \int \nabla_l^2 U(r)[1- \cos(k\cdot r)]g(r) d r,\nonumber\\
\ldots.
\end{eqnarray}
First two parameters were calculated from the molecular dynamics simulation data and the parameters $\Delta_3(k)$ and $\Delta_4(k)$ were taken as fitting values.
Fig. 4 shows the results of theoretical calculations for the scattering intensity $I(k,\omega)$ [Eqs. (5) and (9)] with the detailed balance condition (6) and the experimental resolution function (7) in comparison with the experimental data on inelastic X-ray scattering \cite{Scopigno2006}. It is seen that Eq. (9) yields a good agreement with the experimental data. The results of theoretical calculations reproduce the high-frequency features of the experimental spectra on inelastic X-ray scattering for all values of wave numbers: the correct predictions for the position as well as for the height and damping of the lateral peaks.


\section{Conclusion}

This paper presents the results of the atomic/molecular dynamics simulations of the $Ni_{33}Zr_{67}$ bulk metallic glass at the temperature $T=300K$.  The results indicate on the possibility to perform the correct description of the structural and dynamic properties of the amorphous metallic alloys containing the polyvalent (transition) metals in the framework of pair-wise effective potentials. The simulation results of structural and dynamic characteristics are in a good agreement with the experimental data on neutron diffraction and inelastic X-ray scattering data. The theoretical model developed recently to describe the structural relaxation in liquid metals reproduces the features of the $I(k,\omega)$ spectra for $Ni_{33}Zr_{67}$ metallic glass associated, in particular, with the appearance and disappearance of the high-frequency peaks.

\section*{Acknowledgements}

The authors acknowledge to Dr. T. Scopigno (University of Rome ``La Sapienza'', Italy) and Prof. J.-B. Suck (University of Technology Chemnitz, Germany) for providing the experimental data and discussion of results. This investigation has been partly supported by the Federal Special-Purpose Program ``Cadres'' of the Russian Ministry of Science and Education (project No. 8174).


\end{document}